\documentclass[
    ,final            
    ,a4paper
  ]
  {aipproc}

\layoutstyle{6x9}

\newcommand{\tev}{\,{\rm TeV}}

\newcommand{\mev}{\,{\rm MeV}}

\begin{document}

\title{Nucleon structure in the search for new physics}


\author{R.~D.~Young}{
  address={Special Research Centre for the Subatomic Structure of Matter (CSSM)\\
           and Centre of Excellence in Particle Physics at the Tera-Scale (CoEPP),\\
           School of Chemistry and Physics, University of Adelaide, SA 5005, Australia}
}


\begin{abstract}
  We report on recent results on nucleon structure that are helping
  guide the search for new physics at the precision frontier. Results
  discussed include the electroweak elastic form factors, charge
  symmetry breaking in parton distributions and the strangeness
  content of the nucleon.
\end{abstract}

\maketitle


\section{Introduction}
The Standard Model has been enormously successful at describing
experiments in nuclear and particle physics. The search for new
physical phenomena beyond the Standard Model is primarily driven by
two complementary experimental strategies. The first is to build
high-energy colliders, such as the Large Hadron Collider (LHC) at
CERN, which aim to excite a new form of matter from the vacuum. The
second, more subtle approach is to perform precision measurements at
moderate energies, where an observed discrepancy can signify the
existence such new forms of matter.

The significance of measurements at the precision frontier depends
on careful experimental techniques, in conjunction with robust theoretical predictions of
contributing Standard Model phenomena.  Here we report on some recent
progress in nucleon structure that is contributing to the low-energy
search for new physics.  In particular, the nucleon electroweak
elastic form factors and charge symmetry breaking in parton
distributions both play a significant roles in precision tests of the
weak interaction. In the context of ongoing dark matter searhces,
improved knowledge of the strangeness scalar content of the nucleon is
leading to better constrained predicted cross sections.

%
\section{Quark weak charges}
At low energies, the weak interaction is manifest in the effective
current--current correlators
\begin{equation}
{\cal L}_{PV}=-\frac{G_F}{\sqrt 2}\sum_q\left[C_{1q}\,\overline{e}\gamma^\mu \gamma_5 e\, \overline{q}\gamma_\mu q
+C_{2q}\,\overline{e}\gamma^\mu e\,\overline{q}\gamma_\mu \gamma_5 q
\right]
\end{equation}
where $G_F$ is the weak coupling constant, and the $C_{iq}$ denotes
flavour-dependence of the effective neutral current interaction --- at
tree level they are simply $C_{1(2)q}\sim g_e^{A(V)} g_q^{V(A)}$.  The full
couplings are determined within the Standard Model by combining
precision $Z$-pole measurements \cite{ALEPH:2005ema} with the scale
evolution to the low-energy domain
\cite{Marciano:1983ss,Erler:2003yk}.

\begin{figure}
\resizebox{0.7\textwidth}{!}{\includegraphics{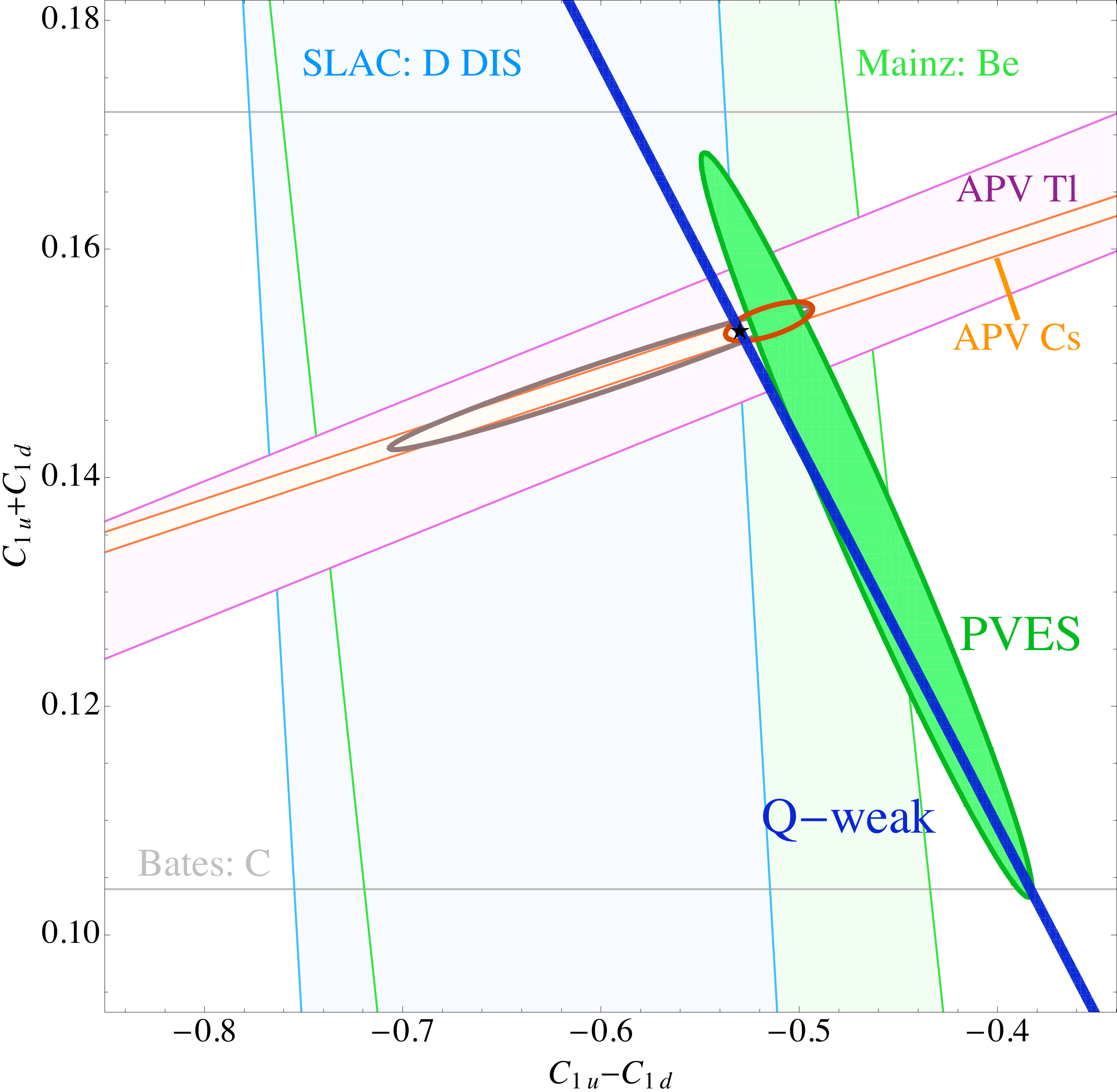}}
\caption{Summary of experimental determination of the weak charges of
  quarks \cite{Young:2007zs}. The Q-weak band indicates the
  anticipated precision of the experiment currently in progress, drawn
  arbitrarily in agreement with the Standard Model.}
\label{fig:c1q}
\end{figure}

Experimental constraints on the weak neutral current at low
energies have been rather limited.  One celebrated result is the
precision measurement of atomic cesium's $6s\to 7s$ transition
polaririzability, and the resulting extraction of the weak nuclear
charge of cesium \cite{Bennett:1999pd}.  The weak charge extraction
depends crucially on the precision calculation of the atomic wave
functions, where the latest theoretical update gives $Q_w^{Cs}\equiv
-376C_{1u}-422C_{1d}=-73.16(29)_{\rm exp}(20)_{\rm th}$
\cite{Porsev:2009pr} --- in complete agreement with the Standard Model
value $-73.15(2)$ \cite{Nakamura:2010zzi}. This agreement with the
Standard Model is depicted by the narrow, almost horizontal (orange) band in
Figure~\ref{fig:c1q}.

The cesium measurement places very restrictive bounds on the form of
parity-violation interactions within new physics scenarios. In terms
of a generic contact interaction describing new physics
\cite{Erler:2003yk}
\begin{equation}
{\cal L}_{PV}^{\rm new}=-\frac{g^2}{4\Lambda^2}\overline{e}\gamma^\mu \gamma_5 e\sum_q h_V^q \overline{q}\gamma_\mu q\,,
\end{equation}
the cesium measurement, at 1-sigma, restricts the magnitude of any new
physics contribution to be less than
\begin{equation}
\frac{g^2}{\Lambda^2}\left(0.67 h_V^u+0.75 h_V^d\right)\sim \left[7{\,\rm TeV}\right]^{-2}.
\end{equation}

The atomic measurements are mostly insensitive to hadronic or nuclear
structure because of the small energy transfers
involved\footnote{Though such effects will become increasingly more
  significant as higher-precision measurements are performed, see
  Ref.~\cite{Brown:2008ib}, for instance}. In electron scattering, the
neutral current can be probed by measuring parity-violating
asymmetries.  Given the typical energy scales involved, the extraction
of the weak interaction parameters also requires knowledge of nucleon
structure. Measurements of this sort date back to the pioneering work
of Prescott {\it et al.}~\cite{Prescott:1979dh} at SLAC, where a
parity-violating asymmetry in deep inelastic scattering was measured
(see the almost vertical band in Figure~\ref{fig:c1q}).

More recently, measurements of the parity-violating {\em elastic}
scattering asymmetries have now been carried out by a number of
experiments, including: SAMPLE at MIT-Bates \cite{Spayde:2003nr}; PVA4 at Mainz
\cite{Maas:2004ta,Maas:2004dh,Baunack:2009gy}; and G0
\cite{Armstrong:2005hs} and HAPPEX
\cite{Aniol:2004hp,Aniol:2005zg,Acha:2006my} at Jefferson Lab. The
principal focus of these programs was the study of the electroweak
form factors of the nucleon, and particularly, the determination of
the strange quark component of these form factors.

\begin{figure}
\resizebox{0.8\textwidth}{!}{\includegraphics{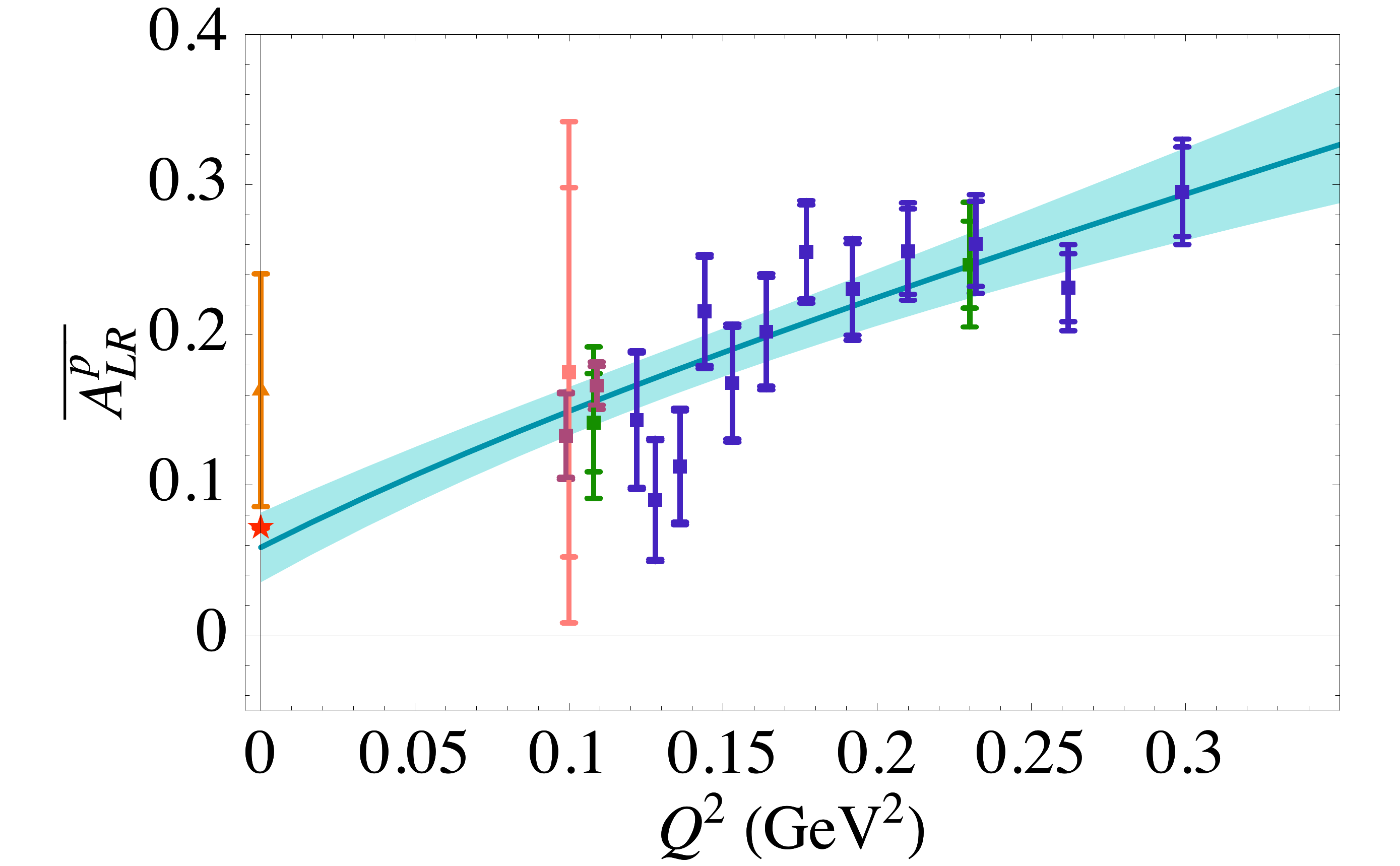}}
\caption{Scaled parity-violating asymmetries measured on a proton
  target, projected onto the forward-angle limit. The normalisation is
  selected such that the $Q^2\to 0$ limit describes the weak charge of
  the proton, $Q_w^p=-2(2C_{1u}+C_{1d})$.}
\label{fig:extrap}
\end{figure}

In addition to the study of the electroweak structure, the kinematic
coverage of these measurements, together with the standard electromagnetic form
factors, provides a reliable extrapolation to the $Q^2\to 0$ limit,
and thereby an extraction of the proton's weak charge
\cite{Young:2007zs}. Figure~\ref{fig:extrap} displays this
extrapolation, where the observed scattering asymmetries (projected
onto the forward limit) are shown. The displayed asymmetry has been
normalised to give the weak charge of the proton at $Q^2=0$.
The slope of the line describes the knowledge of the neutral current
form factors.

The extraction of the proton's weak charge from this modern data
improves on the earlier results by about a factor of 5 --- see the
ellipse in Figure~\ref{fig:c1q}. Following the generic contact
interaction described above, the observed agreement with the Standard
Model sets the characteristic mass scale to above $\sim 2\tev$ (at
1-sigma).

\section{Charge symmetry breaking in parton distributions}
With the improved technology and expertise gained in performing the
precision measurements of the electroweak elastic form factors, there
are now plans to revisit parity-violation in DIS \cite{PVDIS}. This
program is proposing to improve the precision of the early SLAC
measurements of Prescott et al. by roughly an order of magnitude. 

The new Jefferson Lab program is aiming at a sub-1\% measurement of
the PVDIS asymmetry from deuterium. With possible contributions from
supersymmetry, for example, estimated to be as large as $\sim 1\%$
\cite{Kurylov:2003xa}, this program is just at the threshold of a
Standard Model test\footnote{Of course, in conjuction with other
  low-energy measurements, correlations can enhance the significance
  of possible new physics limits.} --- provided Standard Model corrections are
well understood.

One of the potentially largest hadronic corrections to the physics asymmetry
is that arising from charge symmetry violation (CSV). Based on the
phenomenological extraction by Martin et al. \cite{Martin:2003sk}, the
90\% confidence level bounds on CSV lead to $\sim$1.5--2\% variations
in the PVDIS asymmetry \cite{Hobbs:2008mm}. At typical kinematics of
the JLab program, such fluctuations appear to be more significant than
other possible corrections, such as higer twist
\cite{Hobbs:2008mm,Mantry:2010ki} or target-mass corrections
\cite{Hobbs:2011dy}. 

With CSV (potentially) at the scale of $\pm$1.5--2\% of the PVDIS
asymmetry, a precision measurement could provide the best direct
measurement of charge symmetry violation in parton
distributions. While such a measurement would be of great interest for
hadronic physics \cite{Londergan:2006he,londergan:2009kj}, it will
disguise any signature of new physics. Fortunately lattice QCD offers
the opportunity to constrain this hadronic physics independently. In a
recent study, lattice calculations of the hyperon quark momentum
fractions have been used to extract charge symmetry breaking in
nucleon parton distibutions \cite{Horsley:2010th}. These results
suggest CSV in the quark momentum fractions of $\sim 0.20\pm0.06\%$,
corresponding to a $\sim 0.4$--$0.6\%$ correction to the PVDIS
asymmetry. Importantly, the statistical precision represents an order
of magnitude improvement on the bounds reported in
Ref.~\cite{Martin:2003sk}.

With future work to constrain the systematics of the lattice
calculation of CSV and continued theoretical development in
higher-twist and target mass corrections, mentioned above, there is a
strong case that the PVDIS program at JLab will be able to provide an
important new low-energy test of the Standard Model.

We also note that the lattice result of \cite{Horsley:2010th} also
makes an important contribution to the famous NuTeV anomaly
\cite{Zeller:2001hh}. Whereas the original report of a 3-sigma
discrepancy with the Standard Model assumed CSV to be negligible, the value
extracted from the lattice acts to reduce this discrepancy by
1-sigma. The remaining 2-sigma also appear to be naturally described
within the Standard Model as a nuclear medium modification effect
\cite{Cloet:2009qs,Bentz:2009yy}.

\section{Strangeness scalar content}
The strange quark condensate in the nucleon is of particular
significance in the current search for dark matter. The relatively
large coupling of strange quarks to condidate dark matter, combined
with a typically large uncertainty in the strangeness sigma term,
have led to considerable variation in the predicted cross sections for
direct detection measurements \cite{Bottino:1999ei,Ellis:2008hf}.

The traditional method for extracting the strangeness sigma term in
the nucleon, $\sigma_s$, uses the observed hyperon spectrum in
conjuction with the pion-nucleon sigma term
\cite{Gasser:1980sb,Nelson:1987dg}. Even with a perfect extraction of
the light-quark sigma term and best-estimates of higher-order
corrections \cite{Borasoy:1996bx}, this method is limited to an
uncertainty in $\sigma_s$ of $\sim 90\mev$ \cite{Young:2009ps}.

Advances in lattice QCD calculations now provide significantly better
constraint on the strangeness sigma term \cite{Young:2009ps}. There is
general consensus that the strangeness sigma term is on the small side
of early estimates
\cite{Ohki:2008ff,Young:2009zb,Toussaint:2009pz,Ohki:2009mt,MartinCamalich:2010fp,Takeda:2010cw}
--- with a couple recent hints that it may not be quite so small
\cite{Collins:2010gr,Babich:2010at}.

A small strange quark sigma term leads to a dramatic reduction in the
uncertainties of dark matter cross sections \cite{Giedt:2009mr}. For a
range of candidate supersymmetric models of dark matter, the predicted
cross sections are found to be substantially smaller than previously
suggested.

\section{Acknowledgements}
This work was supported by the Australian Research Council.





\bibliographystyle{h-apsrev}

\bibliography{references2}

\begin{thebibliography}{44}
\expandafter\ifx\csname natexlab\endcsname\relax\def\natexlab#1{#1}\fi
\providecommand{\enquote}[1]{``#1''}
\expandafter\ifx\csname url\endcsname\relax
  \def\url#1{\texttt{#1}}\fi
\expandafter\ifx\csname urlprefix\endcsname\relax\def\urlprefix{URL }\fi
\providecommand{\eprint}[2][]{\url{#2}}

\bibitem[Schael et~al.(2006)]{ALEPH:2005ema}
S.~Schael, et~al., \emph{Phys.Rept.} \textbf{427}, 257--454 (2006),
  \eprint{hep-ex/0509008}.

\bibitem[Marciano and Sirlin(1984)]{Marciano:1983ss}
W.~Marciano, and A.~Sirlin, \emph{Phys.Rev.} \textbf{D29}, 75 (1984).

\bibitem[Erler et~al.(2003)]{Erler:2003yk}
J.~Erler, A.~Kurylov, and M.~J. Ramsey-Musolf, \emph{Phys.Rev.} \textbf{D68},
  016006 (2003), \eprint{hep-ph/0302149}.

\bibitem[Bennett and Wieman(1999)]{Bennett:1999pd}
S.~Bennett, and C.~E. Wieman, \emph{Phys.Rev.Lett.} \textbf{82}, 2484--2487
  (1999), \eprint{hep-ex/9903022}.

\bibitem[Porsev et~al.(2009)]{Porsev:2009pr}
S.~Porsev, K.~Beloy, and A.~Derevianko, \emph{Phys.Rev.Lett.} \textbf{102},
  181601 (2009), \eprint{0902.0335}.

\bibitem[Nakamura et~al.(2010)]{Nakamura:2010zzi}
K.~Nakamura, et~al., \emph{J.Phys.G} \textbf{G37}, 075021 (2010).

\bibitem[Brown et~al.(2009)]{Brown:2008ib}
B.~Brown, A.~Derevianko, and V.~Flambaum, \emph{Phys.Rev.} \textbf{C79}, 035501
  (2009), \eprint{0804.4315}.

\bibitem[Prescott et~al.(1979)]{Prescott:1979dh}
C.~Prescott, W.~Atwood, R.~Cottrell, H.~DeStaebler, E.~L. Garwin, et~al.,
  \emph{Phys.Lett.} \textbf{B84}, 524 (1979).

\bibitem[Young et~al.(2007)]{Young:2007zs}
R.~D. Young, R.~D. Carlini, A.~W. Thomas, and J.~Roche, \emph{Phys.Rev.Lett.}
  \textbf{99}, 122003 (2007), \eprint{0704.2618}.

\bibitem[Spayde et~al.(2004)]{Spayde:2003nr}
D.~Spayde, et~al., \emph{Phys.Lett.} \textbf{B583}, 79--86 (2004),
  \eprint{nucl-ex/0312016}.

\bibitem[Maas et~al.(2004)]{Maas:2004ta}
F.~Maas, et~al., \emph{Phys.Rev.Lett.} \textbf{93}, 022002 (2004),
  \eprint{nucl-ex/0401019}.

\bibitem[Maas et~al.(2005)]{Maas:2004dh}
F.~Maas, K.~Aulenbacher, S.~Baunack, L.~Capozza, J.~Diefenbach, et~al.,
  \emph{Phys.Rev.Lett.} \textbf{94}, 152001 (2005), \eprint{nucl-ex/0412030}.

\bibitem[Baunack et~al.(2009)]{Baunack:2009gy}
S.~Baunack, K.~Aulenbacher, D.~Balaguer~Rios, L.~Capozza, J.~Diefenbach,
  et~al., \emph{Phys.Rev.Lett.} \textbf{102}, 151803 (2009),
  \eprint{0903.2733}.

\bibitem[Armstrong et~al.(2005)]{Armstrong:2005hs}
D.~Armstrong, et~al., \emph{Phys.Rev.Lett.} \textbf{95}, 092001 (2005),
  \eprint{nucl-ex/0506021}.

\bibitem[Aniol et~al.(2004)]{Aniol:2004hp}
K.~Aniol, et~al., \emph{Phys.Rev.} \textbf{C69}, 065501 (2004),
  \eprint{nucl-ex/0402004}.

\bibitem[Aniol et~al.(2006)]{Aniol:2005zg}
K.~Aniol, et~al., \emph{Phys.Lett.} \textbf{B635}, 275--279 (2006),
  \eprint{nucl-ex/0506011}.

\bibitem[Acha et~al.(2007)]{Acha:2006my}
A.~Acha, et~al., \emph{Phys.Rev.Lett.} \textbf{98}, 032301 (2007),
  \eprint{nucl-ex/0609002}.

\bibitem[Reimer et~al.(????)]{PVDIS}
P.~Reimer, K.~Paschke, X.~Zheng, et~al., {Jefferson Lab Experiment E1207102}.

\bibitem[Kurylov et~al.(2004)]{Kurylov:2003xa}
A.~Kurylov, M.~Ramsey-Musolf, and S.~Su, \emph{Phys.Lett.} \textbf{B582},
  222--228 (2004), \eprint{hep-ph/0307270}.

\bibitem[Martin et~al.(2004)]{Martin:2003sk}
A.~D. Martin, R.~G. Roberts, W.~J. Stirling, and R.~S. Thorne, \emph{Eur. Phys.
  J.} \textbf{C35}, 325--348 (2004), \eprint{hep-ph/0308087}.

\bibitem[Hobbs and Melnitchouk(2008)]{Hobbs:2008mm}
T.~Hobbs, and W.~Melnitchouk, \emph{Phys.Rev.} \textbf{D77}, 114023 (2008),
  \eprint{0801.4791}.

\bibitem[Mantry et~al.(2010)]{Mantry:2010ki}
S.~Mantry, M.~J. Ramsey-Musolf, and G.~F. Sacco, \emph{Phys.Rev.} \textbf{C82},
  065205 (2010), \eprint{1004.3307}.

\bibitem[Hobbs(2011)]{Hobbs:2011dy}
T.~Hobbs  (2011), \eprint{1102.1106}.

\bibitem[Londergan et~al.(2006)]{Londergan:2006he}
J.~Londergan, D.~Murdock, and A.~W. Thomas, \emph{Phys.Rev.} \textbf{D73},
  076004 (2006), \eprint{hep-ph/0603208}.

\bibitem[Londergan et~al.(2010)]{londergan:2009kj}
J.~T. Londergan, J.~C. Peng, and A.~W. Thomas, \emph{Rev. Mod. Phys.}
  \textbf{82}, 2009--2052 (2010), \eprint{0907.2352}.

\bibitem[Horsley et~al.(2010)]{Horsley:2010th}
R.~Horsley, Y.~Nakamura, D.~Pleiter, P.~Rakow, G.~Schierholz, et~al.  (2010),
  \eprint{1012.0215}.

\bibitem[Zeller et~al.(2002)]{Zeller:2001hh}
G.~P. Zeller, et~al., \emph{Phys. Rev. Lett.} \textbf{88}, 091802 (2002),
  \eprint{hep-ex/0110059}.

\bibitem[Cloet et~al.(2009)]{Cloet:2009qs}
I.~Cloet, W.~Bentz, and A.~Thomas, \emph{Phys.Rev.Lett.} \textbf{102}, 252301
  (2009), \eprint{0901.3559}.

\bibitem[Bentz et~al.(2010)]{Bentz:2009yy}
W.~Bentz, I.~C. Cloet, J.~T. Londergan, and A.~W. Thomas, \emph{Phys. Lett.}
  \textbf{B693}, 462--466 (2010), \eprint{0908.3198}.

\bibitem[Bottino et~al.(2000)]{Bottino:1999ei}
A.~Bottino, F.~Donato, N.~Fornengo, and S.~Scopel, \emph{Astropart.Phys.}
  \textbf{13}, 215--225 (2000), \eprint{hep-ph/9909228}.

\bibitem[Ellis et~al.(2008)]{Ellis:2008hf}
J.~R. Ellis, K.~A. Olive, and C.~Savage, \emph{Phys.Rev.} \textbf{D77}, 065026
  (2008), \eprint{0801.3656}.

\bibitem[Gasser(1981)]{Gasser:1980sb}
J.~Gasser, \emph{Annals Phys.} \textbf{136}, 62 (1981).

\bibitem[Nelson and Kaplan(1987)]{Nelson:1987dg}
A.~E. Nelson, and D.~B. Kaplan, \emph{Phys.Lett.} \textbf{B192}, 193 (1987).

\bibitem[Borasoy and Meissner(1997)]{Borasoy:1996bx}
B.~Borasoy, and U.-G. Meissner, \emph{Annals Phys.} \textbf{254}, 192--232
  (1997), \eprint{hep-ph/9607432}.

\bibitem[Young and Thomas(2010{\natexlab{a}})]{Young:2009ps}
R.~D. Young, and A.~W. Thomas, \emph{Nucl.Phys.} \textbf{A844}, 266C--271C
  (2010{\natexlab{a}}), \eprint{0911.1757}.

\bibitem[Ohki et~al.(2008)]{Ohki:2008ff}
H.~Ohki, H.~Fukaya, S.~Hashimoto, T.~Kaneko, H.~Matsufuru, et~al.,
  \emph{Phys.Rev.} \textbf{D78}, 054502 (2008), \eprint{0806.4744}.

\bibitem[Young and Thomas(2010{\natexlab{b}})]{Young:2009zb}
R.~Young, and A.~Thomas, \emph{Phys.Rev.} \textbf{D81}, 014503
  (2010{\natexlab{b}}), \eprint{0901.3310}.

\bibitem[Toussaint and Freeman(2009)]{Toussaint:2009pz}
D.~Toussaint, and W.~Freeman, \emph{Phys.Rev.Lett.} \textbf{103}, 122002
  (2009), \eprint{0905.2432}.

\bibitem[Ohki et~al.(2009)]{Ohki:2009mt}
H.~Ohki, S.~Aoki, H.~Fukaya, S.~Hashimoto, T.~Kaneko, et~al., \emph{PoS}
  \textbf{LAT2009}, 124 (2009), \eprint{0910.3271}.

\bibitem[Martin~Camalich et~al.(2010)]{MartinCamalich:2010fp}
J.~Martin~Camalich, L.~Geng, and M.~Vicente~Vacas, \emph{Phys.Rev.}
  \textbf{D82}, 074504 (2010), \eprint{1003.1929}.

\bibitem[Takeda et~al.(2010)]{Takeda:2010cw}
K.~Takeda, et~al.  (2010), \eprint{1011.1964}.

\bibitem[Collins et~al.(2010)]{Collins:2010gr}
S.~Collins, G.~Bali, A.~Nobile, A.~Schafer, Y.~Nakamura, and J.~M. Zanotti,
  \emph{PoS} \textbf{LATTICE2010}, 134 (2010), \eprint{1011.2194}.

\bibitem[Babich et~al.(2010)]{Babich:2010at}
R.~Babich, R.~C. Brower, M.~A. Clark, G.~T. Fleming, J.~C. Osborn, et~al.
  (2010), \eprint{1012.0562}.

\bibitem[Giedt et~al.(2009)]{Giedt:2009mr}
J.~Giedt, A.~W. Thomas, and R.~D. Young, \emph{Phys.Rev.Lett.} \textbf{103},
  201802 (2009), \eprint{0907.4177}.

\end{thebibliography}


\end{document}